\documentclass[aps,preprint,showpacs,showkeys,nofootinbib]{revtex4}

\usepackage{graphicx}  %
\usepackage{amsmath}
\usepackage{bm}  %
\usepackage{ulem}
\usepackage{amssymb}
\usepackage{comment}

\newcommand{\bea}{\begin{eqnarray}}
\newcommand{\eea}{\end{eqnarray}}
\newcommand{\beq}{\begin{equation}}
\newcommand{\eeq}{\end{equation}}
\newcommand{\bqa}{\begin{eqnarray}}
\newcommand{\eqa}{\end{eqnarray}}

\def\mqo2{{\!\!\!}}

\begin{document}

\begin{flushright}
{\small
TUM-HEP-1143/18\\
Jun. 7, 2018
}
\end{flushright}

\title{Classical Nonrelativistic Effective Field Theories\\ for a Real  Scalar Field}

\author{Eric Braaten}
\email{braaten@mps.ohio-state.edu}
\affiliation{Department of Physics,
         The Ohio State University, Columbus, OH\ 43210, USA}

\author{Abhishek Mohapatra}
\email{mohapatra.16@buckeyemail.osu.edu}
\affiliation{Department of Physics,
         The Ohio State University, Columbus, OH\ 43210, USA}

\author{Hong Zhang}
\email{hong.zhang@tum.de}
\affiliation{Physik Department T31, Technische Universit\"at M\"unchen,James Franck Stra\ss e 1, D-85748 Garching, Germany}

\date{\today}
%\date{November 2007}

\begin{abstract}
A classical nonrelativistic effective field theory for a real Lorentz-scalar field $\phi$ is most conveniently formulated in terms of a complex scalar field $\psi$. There have been two derivations of effective Lagrangians for the complex field $\psi$ in which terms in the effective potential were determined to order  $(\psi^* \psi)^4$. We point out an error in each of the effective Lagrangians. After correcting the errors, we demonstrate the equivalence of the two effective Lagrangians by verifying that they both reproduce $T$-matrix elements of the relativistic real scalar field theory and by also constructing a redefinition of the complex field $\psi$  that transforms terms in one effective Lagrangian into the corresponding  terms of the other.
\end{abstract}

\smallskip
\maketitle

%%%%%%%%%%%%%%%%%%%%%%%%%%%%
%%%%%%%%%%%%%%%%%%%%%%%%%%%%
%%%%%%%%%%%%%%%%%%%%%%%%%%%%

\section{Introduction}

Scalar fields play an important role in particle physics and in cosmology. A classical scalar field can arise as the vacuum expectation value of  an elementary quantum scalar field or of a scalar composite operator of quantum fields. An example of an elementary scalar field is the Higgs field,  whose  vacuum expectation value generates the spontaneous breaking of the $SU(2) \times U(1)$ symmetry of the Standard Model of particle physics. An example of a scalar field that is not elementary is a pion field, which can be expressed as a composite operator constructed from the quark fields of QCD. A scalar field can arise as the Goldstone mode of a global symmetry. For example, the axion field is a Goldstone mode of a $U(1)$ symmetry of an extension of the Standard Model \cite{Kim:2008hd}. Scalar fields are sometimes used as a simple model for phenomena that might actually be more complicated. For example, the inflation of the early universe  is often described by a single inflaton field.

A Bose-Einstein condensate of spin-0 bosons can also  be described by a classical scalar field. A real Lorentz scalar field $\phi(x)$ has quanta that are identical spin-0 bosons. The bosons are their own antiparticles, but if the annihilation reaction is suppressed, we can consider many-body systems of the bosons. A Bose-Einstein condensate is a system with a large number  of bosons in the same quantum state. The condensate  can be described by a classical complex mean field $\psi(x)$  that is proportional to the common wavefunction of the bosons and is normalized so that $\psi^* \psi$ is the number density of bosons. The mean field can be interpreted as the vacuum expectation value of a nonrelativistic quantum field $\psi(x)$. It is therefore  plausible that the nonrelativistic reduction of  the relativistic quantum field theory for a real scalar field $\phi(x)$ gives a nonrelativistic quantum field theory with a complex scalar field $\psi(x)$.

It is somewhat counterintuitive that the nonrelativistic reduction of  the  real scalar field $\phi(x)$ could be a complex scalar field $\psi(x)$, because its real and imaginary components are two real scalar fields. Despite the mismatch in the number of fields, they describe the same number of degrees of freedom. The resolution of this puzzle is that the relativistic Lagrangian has two time derivatives of $\phi(x)$, while the nonrelativistic Lagrangian has only single time derivatives of $\psi(x)$. The number of propagating degrees of freedom is equal to  the number of real fields if the Lagrangian is second order in their time derivatives, and it is equal to half the number of real fields if the Lagrangian is first order in their time derivatives.

A Bose-Einstein condensate of identical spin-0 bosons of mass $m$ whose number density varies sufficiently slowly in space and time can be described by a classical mean field  $\psi(x)$ with only wavelengths large compared to $2\pi/m$ and with only angular frequencies small compared to $m$. The connection between the real field $\phi(x)$ and the complex field $\psi(x)$ can be expressed naively as
%===========================
\begin{equation}
\phi(\bm{r},t)  = \frac{1}{\sqrt{2m}}
\left[ \psi(\bm{r},t) e^{-imt} + \psi^*(\bm{r},t) e^{+imt} \right].
\label{eq:theta-psi}
\end{equation}
%===========================
A naive effective Lagrangian for $\psi$ can be obtained by substituting the expression in Eq.~\eqref{eq:theta-psi} into the Lagrangian for $\phi$ and omitting all terms with  frequencies of order $m$. This effective Lagrangian may be adequate if the particles have no interactions. However if there are interactions, the field $\phi(\bm{r},t)$ necessarily has frequencies that are harmonics of the fundamental angular frequency $m$. The higher harmonics and other angular frequencies  whose difference from $m$ is of order $m$ arise either from particles that are highly virtual or from relativistic particles that that are on-shell. In either case, their effects  on nonrelativistic particles can be localized to within a distance $1/m$ \cite{Braaten:2016sja}. Thus it should be possible to develop a classical nonrelativistic effective field theory (CNREFT) with a local Lagrangian that reproduces the classical relativistic field theory with systematically improvable accuracy at long wavelengths and at angular frequencies close to $m$.

An interesting application of CNREFT is to axions, which are a well-motivated candidate for the particles that make up the dark matter of the universe \cite{Kim:2008hd}. The axions produced in the early universe are highly nonrelativistic and have huge occupation numbers \cite{Davis:1986xc,Harari:1987ht,Preskill:1982cy,Abbott:1982af,Dine:1982ah}. Gravitational interactions may be able to thermalize the axions into a Bose-Einstein condensate (BEC) \cite{Sikivie:2009qn}. Some aspects of the axion BEC can be described more simply using a CNREFT with complex field $\psi(x)$ than with the relativistic field theory with real field $\phi(x)$.

Another interesting application of CNREFT is to {\it oscillons}. Oscillons are approximately localized solutions of the classical field equations for a real scalar field that remain approximately periodic for a very large number of oscillation periods. They were first discovered by Bogolubsky and Makhankov \cite{BM:1976}. They were subsequently studied by Gleiser, who proposed the name ``oscillons'' \cite{Gleiser:1993pt}. In a real scalar field theory with an interaction potential $V(\phi)$ that allows oscillons, a spherically symmetric initial configuration that is localized in a region much larger than the Compton wavelength $2\pi/m$ will typically relax to an oscillon configuration by radiating away a significant fraction of its initial energy. After remaining apparently stable for a very large number of oscillations with angular frequency near $m$, the oscillon becomes unstable and  it rather quickly dissipates into outgoing waves. Assuming the oscillon configuration has only long wavelengths much larger than $2\pi/m$, it can be described more simply by CNREFT with the slowly varying complex field $\psi(\bm{r},t)$ instead of the relativistic field theory with  the rapidly oscillating real field $\phi(\bm{r},t)$.

A CNREFT for a real scalar field $\phi(x)$ with interactions given by a potential $V(\phi)$ was first constructed explicitly by us in Ref.~\cite{Braaten:2016kzc} using effective field theory methods. The effective field theory for the complex field $\psi$ was called {\it axion EFT}, because it was applied to the axion field whose potential $V(\phi)$ is a periodic function of $\phi$. The effective Lagrangian has the conventional form for a nonrelativistic field theory, in which the only term that depends on the time derivative of $\psi$ is a single term with one time derivative. The real part of the effective potential $V_\mathrm{eff}(\psi^*\psi)$ for the complex field was calculated to fifth order in $\psi^*\psi$ for a general potential $V(\phi)$ that is invariant under the  $Z_2$ symmetry $\phi(x) \to - \phi(x)$ and has a minimum at $\phi=0$. In Ref.~\cite{Braaten:2016dlp}, the imaginary part of the effective potential was calculated to sixth order in $\psi^*\psi$. If the power series for the potential $V(\phi)$ has a finite radius of convergence, the power series for  the effective potential $V_\mathrm{eff}(\psi^*\psi)$ also has a finite radius of convergence. In Ref.~\cite{Braaten:2016kzc}, a resummation method was developed to calculate $V_\mathrm{eff}(\psi^*\psi)$ beyond the radius of convergence of its power series.

There have been several subsequent efforts to derive a CNREFT for a relativistic field theory with a real scalar field. In Ref.~\cite{Mukaida:2016hwd}, Mukaida, Takimoto and Yamada constructed the nonrelativistic effective Lagrangian for $\psi(x)$ by integrating out relativistic fluctuations of  $\phi(x)$. Their effective Lagrangian for $\psi$ is completely different from that in Ref.~\cite{Braaten:2016kzc}, having a term with two time derivatives of $\psi$ and an effective potential that is different beginning with the $(\psi^*\psi)^3$ term. In Ref.~\cite{Namjoo:2017nia}, Namjoo, Guth, and Kaiser discovered an exact transformation  between the real scalar field $\phi(x)$ and a complex field $\psi(x)$ that is a generalization of the naive relation in Eq.~\eqref{eq:theta-psi} which is nonlocal in space. They used the transformation to verify the $(\psi^*\psi)^3$ term in the effective potential $V_\mathrm{eff}(\psi^*\psi)$ in Ref.~\cite{Braaten:2016kzc}. In Ref.~\cite{Eby:2017teq}, Eby, Suranyi, and Wijewardhana developed a method that gives a sequence of improvements to equations for a complex field $\psi(x)$  with harmonic time dependence.

In this paper,  we elucidate the relations between the classical effective field theories in Refs.~\cite{Braaten:2016kzc}, \cite{Mukaida:2016hwd}, \cite{Namjoo:2017nia}, and \cite{Eby:2017teq}.  All four papers give results that depend on the $(\psi^*\psi)^3$ term in the effective potential, and they are all consistent at that order. The only papers in which $(\psi^*\psi)^4$ terms in the effective potential were determined were Refs.~\cite{Braaten:2016kzc} and \cite{Mukaida:2016hwd}. We identify an error in the  effective Lagrangian in both papers. In Ref.~\cite{Braaten:2016kzc}, the failure to take into account gradient interaction terms in the effective Lagrangian led to an error in the coefficient of the $(\psi^*\psi)^4$ term in the effective potential. In Ref.~\cite{Mukaida:2016hwd}, the error in the  effective Lagrangian is the omission of an interaction term with time derivatives. After correcting the errors, we demonstrate the equivalence of the effective Lagrangians in Refs.~\cite{Braaten:2016kzc} and \cite{Mukaida:2016hwd} by verifying that they both reproduce $T$-matrix elements of the relativistic real scalar field theory and by constructing a redefinition of the complex field $\psi$  that transforms terms in one effective Lagrangian into the corresponding terms  of the other.

We begin in Section~\ref{sec:Relativistic} by defining the relativistic field theory of the real scalar field. In Section~\ref{sec:BMZ}, we describe the CNREFT we constructed in Ref.~\cite{Braaten:2016kzc} and we point out the error in the calculation of the effective potential. In Section~\ref{sec:Namjoo}, we describe the exact transformation between the real scalar field $\phi$ and a complex field $\psi$  discovered by Namjoo et al.\ in Ref.~\cite{Namjoo:2017nia}. In Section~\ref{sec:MTY}, we describe the  CNREFT constructed by Mukaida et al.\  in Ref.~\cite{Mukaida:2016hwd} and we point out the time-derivative interaction term that was omitted from their effective Lagrangian. After correcting the errors in the effective Lagrangians in Refs.~\cite{Braaten:2016kzc} and \cite{Mukaida:2016hwd}, we demonstrate their equivalence in two different ways. In Section~\ref{sec:ESW}, we show that the equations for the field $\psi(x)$ with harmonic time dependence derived by Eby et al.\ in Ref.~\cite{Eby:2017teq} are consistent with the field equation from the effective Lagrangian in Ref.~\cite{Mukaida:2016hwd}. We summarize our results in Section~\ref{sec:Summary}.

%%%%%%%%%%%%%%%%%%%%%%%%%%%%
%%%%%%%%%%%%%%%%%%%%%%%%%%%%
%%%%%%%%%%%%%%%%%%%%%%%%%%%%
\section{Relativistic  field theory} \label{sec:Relativistic}

The Lorentz-invariant Lagrangian for a real scalar field $\phi(x)$ is
%===========================
\begin{equation}
\mathcal{L} =
\frac{1}{2}  \partial^\mu\phi \partial_\mu\phi - \frac12 m^2 \phi^2- V(\phi) ,
\label{eq:L-real}
\end{equation}
%===========================
where  $m$ is a positive parameter  and $V(\phi)$ is the interaction potential energy density. Quantization of the noninteracting field theory with $V=0$ produces quanta that are spin-0 bosons with mass $m$ and with the relativistic energy-momentum relation
%===========================
\begin{equation}
E = \sqrt{m^2 +\bm{p}^2}.
\label{eq:E-p}
\end{equation}
%===========================
The classical equations of motion are 
%===========================
\begin{equation}
\ddot \phi = \bm{\nabla}^2 \phi  - m^2 \phi - V'(\phi) .
\label{eq:thetaEq}
\end{equation}
%===========================
In the classical theory, the energy-momentum relation in  Eq.~\eqref{eq:E-p} gives the dispersion relation for low-amplitude waves with angular frequency $E$ and wavevector $\bm{p}$.

We assume that $\phi=0$ is a minimum of $V(\phi)$. We choose to consider field theories with the $Z_2$ symmetry $\phi \to -\phi$, which requires $V$ to be an even function of $\phi$. It  can therefore be expanded in powers of $\phi^2$:
%===========================
\begin{equation}
V(\phi) = 
m^2 f^2 \sum_{n=2}^{\infty} \frac{\lambda_{2n}}{(2n)!} \big(\phi^2/f^2 \big)^n,
\label{eq:V-series}
\end{equation}
%===========================
where  the coefficients  $\lambda_{2n}$ are dimensionless coupling constants. The positive parameter $f$ provides a loop-counting device:  if $m/f$ is small, loop diagrams are suppressed by a factor of $m^2/f^2$ for every loop. Our primary interest is in the classical field theory, so we assume $m\ll f$, and we usually keep only the leading term in the expansion in powers of $m/f$.

For a fundamental quantum field theory, renormalizability requires the series for the effective potential in Eq.~\eqref{eq:V-series} to terminate at the power $\phi^4$. This requirement is unnecessary for  a classical field theory or for an effective quantum field theory. If the series terminates with some maximum even power $\phi^{2N}$, classical stability requires its coupling constant $\lambda_{2N}$ to be positive. An important example in which the series does not terminate is the  {\it sine-Gordon model},  in which the potential energy density is
%===========================
\begin{equation}
 \frac12 m^2 \phi^2 + V(\phi) =
m^2 f^2 \big[1- \cos (\phi/f)\big].
\label{eq:V-sineGordon}
\end{equation}
%===========================
This model is often used as an effective field theory for the {\it axion}.

Oscillons  were first discovered by Bogolubsky and Makhankov in a real scalar field theory with a $\phi^4$ interaction potential with a $Z_2$ symmetry but with $m^2<0$ \cite{BM:1976}. Our assumption that $V(\phi)$ has a minimum at $\phi=0$ excludes this symmetric double-well potential. Oscillons require an interaction potential such that $m^2+V''(\phi)<0$ for some region of $\phi$. They arise in the sine-Gordon model \cite{Bogolubsky:1977}. They also arise in a model whose interaction potential has $\phi^4$ and $\phi^6$ terms with $\lambda_4<0$ and $\lambda_6>0$.

%%%%%%%%%%%%%%%%%%%%%%%%%%%%
%%%%%%%%%%%%%%%%%%%%%%%%%%%%
%%%%%%%%%%%%%%%%%%%%%%%%%%%%
\section{Nonrelativistic effective field theory}
\label{sec:BMZ}

Identical relativistic spin-0 bosons can be described by a field theory with a real  Lorentz-scalar field $\phi(x)$. However nonrelativistic particles can be described more simply by a  nonrelativistic field theory with a complex scalar field $\psi(\bm{r}, t)$. A complex field can be introduced naively by expressing the real scalar field in the form in Eq.~\eqref{eq:theta-psi}, where $\psi(\bm{r}, t)$ has only wavelengths large compared to $1/m$ and only frequencies small compared to $m$. A naive effective Lagrangian for  $\psi$ can be obtained by inserting the expression for $\phi$ in Eq.~\eqref{eq:theta-psi} into the Lagrangian in Eq.~\eqref{eq:L-real}, dropping terms with a rapidly changing phase factor of the form $\exp(i n mt)$, where $n$ is a nonzero integer, and dropping terms in which the time derivative $\dot \psi$ is divided by $m$. The resulting Lagrangian has the form
%===========================
\begin{equation}
{\cal L}_\mathrm{naive} = 
\frac{i}{2}\left( \psi^* \dot \psi - \dot \psi^*\psi \right) 
- \frac{1}{2m} \bm{\nabla} \psi^* \!\cdot\! \bm{\nabla} \psi
- V_\mathrm{naive}(\psi^* \psi).
\label{eq:Lnaive}
\end{equation}
%===========================
If the potential energy density $V(\phi)$ for the real scalar field has the power series in Eq.~\eqref{eq:V-series}, the power series  for the naive effective potential is 
%===========================
\begin{equation}
V_\mathrm{naive}(\psi^* \psi) = 
m^2 f^2 \sum_{j=2}^{\infty} \frac{\lambda_{2j}}{(j!)^2} \left(\frac{\psi^* \psi}{2m f^2} \right)^j.
\label{eq:Veff-naive}
\end{equation}
%===========================

A nonrelativistic effective field theory that provides a systematically improvable approximation to the relativistic field theory in the nonrelativistic region can be obtained rigorously by using effective field theory methods. We developed a classical nonrelativistic effective field theory (CNREFT) for a real Lorentz-scalar field in Ref.~\cite{Braaten:2016kzc}.  We called it {\it axion EFT}, because we applied it to the axion field. The effective Lagrangian has the form
%===========================
\begin{equation}
{\cal L}_\mathrm{eff} = 
\tfrac12 i \left( \psi^* \dot \psi - \dot \psi^*\psi \right) - {\cal H}_\mathrm{eff},
\label{Leff-psi}
\end{equation}
%===========================
where ${\cal H}_\mathrm{eff}$ depends only on $\psi$ and $\psi^*$ and their spatial derivatives. The effective Lagrangian has a $U(1)$ symmetry in which the field $\psi(x)$ is multiplied by a phase. This ensures that every term has equal numbers of factors of $\psi$ and $\psi^*$. We refer to a term with $n$ factors of $\psi$ and $n$ factors of $\psi^*$ as an $n$-body term. The only term that depends on the time derivative $\dot \psi$ is the 1-body term that appears explicitly in Eq.~\eqref{Leff-psi}. The effective Hamiltonian density has the form
%===========================
\begin{equation}
\mathcal{H}_\mathrm{eff} = 
\mathcal{T}_\mathrm{eff} + \big[ V_\mathrm{eff} + W_\mathrm{eff}  \big]
-i  \big[ X_\mathrm{eff} + Y_\mathrm{eff}  \big] .
\label{Heff-psi}
\end{equation}
%===========================
The kinetic energy density  $\mathcal{T}_\mathrm{eff}$ consists of 1-body terms with at least two gradients. The remaining terms are interaction terms that are 2-body and higher, and they have been separated into real and imaginary parts. The effective potential  $V_\mathrm{eff}- i X_\mathrm{eff}$ is a function of $\psi^*\psi$. The terms in  $W_\mathrm{eff} - i Y_\mathrm{eff}$ have two or more gradients acting on $\psi$ and $\psi^*$.

The kinetic energy density is
%===========================
\begin{equation}
\mathcal{T}_\mathrm{eff} = 
\frac{1}{2m} \bm{\nabla} \psi^* \!\cdot\! \bm{\nabla} \psi - \frac{1}{8m^3}\bm{\nabla}^2 \psi^* \bm{\nabla}^2 \psi
+ \frac{1}{16 m^5} \bm{\nabla}(\bm{\nabla}^2 \psi^*)  \!\cdot\! \bm{\nabla}(\bm{\nabla}^2 \psi) + \ldots.
\label{Teff-psi}
\end{equation}
%===========================
Only the $\bm{\nabla} \psi^* \cdot \bm{\nabla} \psi$ term was given explicitly in Ref.~\cite{Braaten:2016kzc}. The terms with additional gradients are needed to reproduce the relativistic energy-momentum relation in  Eq.~\eqref{eq:E-p} to higher orders in $\bm{p}^2/m^2$.

The real part of the effective potential can be expanded in powers of $\psi^* \psi$ beginning at order $(\psi^*\psi)^2$:
%===========================
\begin{equation}
V_\mathrm{eff}(\psi^* \psi) = 
m^2 f^2 \sum_{j=2}^{\infty} \frac{v_j}{(j!)^2} \left(\frac{\psi^* \psi}{2m f^2} \right)^j.
\label{eq:Veff-series}
\end{equation}
%===========================

In Ref.~\cite{Braaten:2016kzc}, the mass energy term $m\psi^*\psi$ was included in the effective potential. The equations of motion then imply that $\psi(\bm{r},t)$ has  a multiplicative factor $e^{i mt}$, so it has large frequencies of order $m$. In the effective field theory, it is more appropriate to eliminate the scale $m$, so that $\psi(\bm{r},t)$ has only frequencies much smaller than $m$. We therefore omit the term $m\psi^*\psi$ from the effective potential. In the classical effective field theory, we keep only the leading terms in the  expansions of the real dimensionless coefficients $v_j$ in powers of $m^2/f^2$. The coefficient  $v_j$ can be determined by calculating the tree-level $j \to j$  $T$-matrix element in the relativistic theory  and matching it with the tree-level $T$-matrix element in  the nonrelativistic EFT  in the limit where the 3-momenta go to zero \cite{Braaten:2016kzc}. The coefficient  $v_j$ is determined by  the coefficients $\lambda_{2n}$ in the interaction potential  $V(\phi)$ for the real scalar field in Eq.~\eqref{eq:V-series}  with $n \le j$. The first three coefficients are
%===========================
\begin{subequations}
\begin{eqnarray}
v_2 &=& \lambda_4,
\label{v2}
\\
v_3 &=&  \lambda_6  -\frac{17}{8} \lambda_4^2,
\label{v3}
\\
v_4 &=&  \lambda_8 - 11 \lambda_4 \lambda_6 + \frac{125}{8} \lambda_4^3.
\label{v4}
\end{eqnarray}
\label{v234}%
\end{subequations}
%===========================
The coefficient $v_2$ is the same as in the naive effective potential in Eq.~\eqref{eq:Veff-naive}. The coefficients $v_3$, $v_4$, and $v_5$ were calculated in Ref.~\cite{Braaten:2016kzc}, but there were errors in the coefficients $v_4$ and $v_5$. The error in $v_4$ has been corrected in Eq.~\eqref{v4}.

The error made in Ref.~\cite{Braaten:2016kzc} was assuming that interaction terms in $W_\mathrm{eff}$, which depend on gradients of $\psi$ and $\psi^*$, can be ignored in the matching calculations of $T$-matrix elements  in the limit where the 3-momenta go to zero. In that limit, the Feynman rule for the $j \to j$ vertex  in CNREFT is $-i v_j m^2 f^2/(2mf^2)^j$. These vertices were used in the matching calculation in Ref.~\cite{Braaten:2016kzc}. The resulting expression for $v_4$ is given by Eq.~\eqref{v4}, with the coefficient 125/8 of $\lambda_4^3$ replaced by 49/4. When the gradient  interaction terms in $W_\mathrm{eff}$ are taken into account, the vertices become momentum dependent \cite{HYZ:1801}. For example, at 2$^\mathrm{nd}$ order in the 3-momenta, the  Feynman rule for the $2 \to 2$ vertex with incoming 3-momenta $\bm{p}_1$ and $\bm{p}_2$ and outgoing  3-momenta $\bm{p}_1'$ and $\bm{p}_2'$ is $-i (\lambda_4/2f^2) \big[1 - (\bm{p}_1^2 + \bm{p}_2^2 + \bm{p}_1'{^2} + \bm{p}_2'{^2})/(4m^2) \big]$. To obtain the correct coefficient $v_4$ in Eq.~\eqref{v4}, it is necessary to keep terms in the $2 \to 2$ vertex up to 4$^\mathrm{th}$ order in the 3-momenta and terms in the $3 \to 3$ vertex up to 2$^\mathrm{nd}$ order in the 3-momenta, even when matching the $4 \to 4$ $T$-matrix element in the limit where the 3-momenta go to zero.

The imaginary part $X_\mathrm{eff}$  of the effective potential  is suppressed relative to $V_\mathrm{eff}$ by a factor of $m^2/f^2$. It can be expanded in powers of $(\psi^* \psi)^2$ beginning at order $(\psi^*\psi)^4$ \cite{Braaten:2016dlp}:
%===========================
\begin{equation}
X_\mathrm{eff}(\psi^* \psi) = 
 m^4 \sum_{j=2}^{\infty} \frac{x_{2j}}{[(2j)!]^2} \left(\frac{\psi^* \psi}{2m f^2} \right)^{2j}.
\label{eq:Xeff-series}
\end{equation}
%===========================
The coefficient $x_{2j}$ can be determined by calculating the tree-level $2j \to2$ $T$-matrix element for bosons in the relativistic theory  and matching its square with the imaginary part of the tree-level $2j \to2j$  $T$-matrix element in the nonrelativistic EFT in the limit where the incoming 3-momenta go to zero \cite{Braaten:2016dlp}. The coefficients $x_4$ and $x_6$  were calculated in Ref.~\cite{Braaten:2016dlp}. If the effective Hamiltonian was Hermitian, the $U(1)$ symmetry of the effective Lagrangian would imply  conservation of the particle number: 
%===========================
\begin{equation}
N = \int\!\! d^3r\,  \psi^* \psi .
\label{eq:Naxion}
\end{equation}
%===========================
However the Hamiltonian density in Eq.~\eqref{Heff-psi} includes the anti-hermitian terms $-iX_\mathrm{eff}$ and $-iY_\mathrm{eff}$. The rate of decrease in the particle number is given by
%===========================
\begin{equation}
\frac{d\ }{dt}N = - 2 \int\!\! d^3r\, \big[   X_\mathrm{eff}'(\psi^* \psi) \, \psi^* \psi  + \ldots \big],
\label{dNaxion/dt}
\end{equation}
%===========================
where $X_\mathrm{eff}'$ denotes the derivative of $X_\mathrm{eff}$ with respect to its argument $\psi^* \psi$. The additional terms not shown explicitly come from the interaction terms in $Y_\mathrm{eff}$, which have gradients of $\psi$ and $\psi^*$.

%%%%%%%%%%%%%%%%%%%%%%%%%%%%
%%%%%%%%%%%%%%%%%%%%%%%%%%%%
%%%%%%%%%%%%%%%%%%%%%%%%%%%%
\section{Nonlocal field  Transformation}\label{sec:Namjoo}

In Ref.~\cite{Namjoo:2017nia}, Namjoo, Guth, and Kaiser discovered an exact transformation between a relativistic real scalar field $\phi(x)$ with the Lagrangian in Eq.~\eqref{eq:L-real} and a complex field $\psi(x)$.\footnote{This transformation has appeared previously as a problem in a textbook \cite{Gross1993}.} The Lagrangian for $\psi$ has the form in Eq.~\eqref{Leff-psi}, with a 1-body term that is first order in the time derivative $\dot \psi$ and a nonlocal Hamiltonian density $\mathcal{H}(\bm{r},t)$ that depends explicitly on the time $t$ and depends on the field $\psi(\bm{r}',t)$ everywhere in space. Their exact relation between $\phi(\bm{r},t)$ and $\psi(\bm{r},t)$ is 
%===========================
\begin{equation}
\phi(\bm{r},t) = \frac{1}{\sqrt{2m}} \left( 1-\frac{\bm{\nabla}^2}{m^2} \right)^{-1/4}
\Big( \psi(\bm{r},t) \, e^{-i m t} + \psi^*(\bm{r},t) \, e^{+i m t} \Big).
\label{eq:theta-psiNGK}
\end{equation}
%===========================
The relation between the real fields $\phi$ and $\dot \phi$ and the complex fields  $\psi$ and $\psi^\dagger$ is a canonical transformation. Despite the relation being nonlocal in space, the local equal-time commutation relations between $\phi$ and its canonical momentum imply the canonical local equal-time commutation relations between $\psi$ and $\psi^*$.

If one considers complex fields $\psi(\bm{r},t)$ with gradients small compared to $2\pi/m$ and frequencies small compared to $m$, the Hamiltonian density for $\psi$ can be expanded in terms of local operators with no explicit time dependence and no time derivatives \cite{Namjoo:2017nia}. The resulting Lagrangian for $\psi$ has the  same form as the effective Lagrangian for the CNREFT in Eq.~\eqref{Leff-psi}. Namjoo et al.\ used operator methods based on their exact  transformation to derive terms in the effective Lagrangian \cite{Namjoo:2017nia}. They verified the results in Eq.~\eqref{v2} and \eqref{v3} for the coefficients $v_2$ and $v_3$ in the real part of the effective potential. They also calculated an interaction term in the effective Lagrangian that depends on gradients of the field:\footnote{We have corrected an apparent typographical error in the power of $m$ in the coefficient.}
%===========================
\begin{equation}
W_\mathrm{eff} = 
\frac{\lambda_4}{16m} \left(\frac{\psi^*\psi}{2m f^2}\right) 
\big(\psi^*\bm{\nabla}^2\psi+\bm{\nabla}^2\psi^* \psi\big)+ \ldots.
\label{eq:Weff-NGK}
\end{equation}
%===========================

In Ref.~\cite{Schiappacasse:2017ham}, Schiappacasse and Hertzberg claimed that for the sine-Gordon model, whose interaction potential is given in Eq.~\eqref{eq:V-sineGordon}, the nonrelativistic EFT with complex field $\psi$ can only be applied in the dilute region where $\psi^*\psi \ll 2 \pi^2 m f^2$, which corresponds to $|\phi| \ll 2 \pi f$. Their primary argument was that the shift symmetry $\phi \to \phi + 2 \pi f$ of the scalar field associated with the periodicity of  $V(\phi)$ is not evident in the Lagrangian for the nonrelativistic EFT. Their argument is convincingly refuted by the discovery of the exact transformation between $\phi$ and $\psi$ by Namjoo et al.\ \cite{Namjoo:2017nia}. This relation makes it clear that the shift symmetry of  $\phi$ is not broken when the relativistic field theory is expressed in terms of the complex field $\psi$, but is only hidden.  It suggests that the shift symmetry may also be only hidden in the CNREFT.

%%%%%%%%%%%%%%%%%%%%%%%%%%%%
%%%%%%%%%%%%%%%%%%%%%%%%%%%%
%%%%%%%%%%%%%%%%%%%%%%%%%%%%
\section{Integrating out relativistic fluctuations}
\label{sec:MTY}

In Ref.~\cite{Mukaida:2016hwd}, Mukaida, Takimoto and Yamada developed a CNREFT in order to study oscillons in a relativistic field theory with a real scalar field. Their effective Lagrangian  is very different from the effective Lagrangian in Ref.~\cite{Braaten:2016kzc}. After correcting an error in the effective Lagrangian of Ref.~\cite{Mukaida:2016hwd}, we demonstrate its equivalence with the corrected effective Lagrangian of Ref.~\cite{Braaten:2016kzc}. We show that the two effective Lagrangians give the same $T$-matrix elements, and we show that they differ by a redefinition of the complex field $\psi$.

%%%%%%%%%%%%%%%%%%%%%%%%%%%%
%%%%%%%%%%%%%%%%%%%%%%%%%%%%
\subsection{MTY Effective Lagrangian}
\label{sec:Leff-MTY}

In Ref.~\cite{Mukaida:2016hwd}, Mukaida et al.\  considered a real scalar field $\phi$ with an interaction potential $V(\phi)$ that has a minimum at $\phi=0$ and that has a term $g_3\phi^3/3$ in addition to even powers of $\phi$. We choose to consider only field theories with the $Z_2$ symmetry $\phi \to - \phi$, so we set $g_3=0$.

In Ref.~\cite{Mukaida:2016hwd}, Mukaida et al.\  derived a nonrelativistic effective  Lagrangian for a complex scalar field $\psi$ by integrating out relativistic  fluctuations of the real field $\phi$. We refer to the CNREFT defined by this Lagrangian as the {\it MTY effective  theory}. We choose to rescale their complex field by a factor of $\sqrt{2/m}$ so that the $\bm{\nabla}\psi^*\cdot\!\bm{\nabla}\psi$ term in the Lagrangian has the standard coefficient $-1/2m$. 
The real part of the MTY effective Lagrangian has the  form
%===========================
\begin{equation}
\mathrm{Re}[\mathcal{L}_\mathrm{MTY}] =\frac{i }{2}  \left( \psi^* \dot \psi - \dot \psi^*\psi \right)
+ \frac{1}{2m} \dot\psi^* \dot \psi 
- \frac{1}{2m} \bm{\nabla} \psi^* \!\cdot\! \bm{\nabla} \psi 
-  V_\mathrm{MTY} - W_\mathrm{MTY} .
\label{eq:LMTY_rescaled}
\end{equation}
%===========================
No imaginary terms in the MTY effective Lagrangian were determined  in Ref.~\cite{Mukaida:2016hwd}. The real part of $\mathcal{L}_\mathrm{MTY}$ has 1-body terms with both one and two time derivatives. The only other 1-body term is the $\bm{\nabla} \psi^* \!\cdot\! \bm{\nabla} \psi$ term.  The interaction terms consist of the effective potential  $V_\mathrm{MTY}$, which is a function of $\psi^*\psi$, and $W_{\mathrm{MTY}}$, which consists of terms with gradients or time derivatives of $\psi$ and $\psi^*$. The effective potential can be expressed as a power series in $\psi^*\psi$:
%===========================
\begin{equation}
V_\mathrm{MTY}(\psi^* \psi) = 
m^2 f^2 \sum_{j=2}^{\infty} \frac{z_j}{(j!)^2} \left(\frac{\psi^* \psi}{2m f^2} \right)^j.
\label{eq:VMTY-series}
\end{equation}
%===========================
If the  potential $V(\phi)$ for the real scalar field has the power series in Eq.~\eqref{eq:V-series}, the coefficients of the first few terms are
%===========================
\begin{subequations}
\begin{eqnarray}
z_2 &=& \lambda_4,
\label{z2}
\\
z_3 &=&  \lambda_6 + \frac{1}{8} \lambda_4^2,
\label{z3}
\\
z_4 &=&  \lambda_8 + x \lambda_4 \lambda_6 + \frac{1}{4}\lambda_4^3.
\label{z4}
\end{eqnarray}
\label{z234}%
\end{subequations}
%===========================
The coefficients $z_2$ and $z_3$ and the $\lambda_4^3$ term in $z_4$ were given in Ref.~\cite{Mukaida:2016hwd}. The $\lambda_8$ term in $z_4$ can be deduced from the naive effective potential in Eq.~\eqref{eq:Veff-naive}. The coefficient $x$ of  $\lambda_4 \lambda_6$ in $z_4$ was not given in Ref.~\cite{Mukaida:2016hwd}. The  terms in $W_\mathrm{MTY}$ that were determined  in Ref.~\cite{Mukaida:2016hwd} reduce in the case $g_3=0$ to a single 3-body interaction term:
%===========================
\begin{equation}
W_\mathrm{MTY}=  
\frac{1}{512m}\lambda_4^2 \left(\frac{\psi^* \psi}{2 m f^2} \right)^2  \bm{\nabla}\psi^*\!\cdot\!\bm{\nabla}\psi
+ \ldots.
\label{eq:WMTY-gradients}
\end{equation}
%===========================

The MTY effective Lagrangian in Ref.~\cite{Mukaida:2016hwd} differs from the effective Lagrangian in Ref.~\cite{Braaten:2016kzc} in many ways. The effective Lagrangian in Eq.~\eqref{eq:LMTY_rescaled} has an additional $\dot{\psi}^*\dot{\psi}$  term with two time derivatives that does not appear in Eq.~\eqref{Leff-psi}. The only 1-body  term with gradients in Eq.~\eqref{eq:LMTY_rescaled} is the $ \bm{\nabla} \psi^* \!\cdot\! \bm{\nabla} \psi $ term, while there are infinitely many terms in the kinetic energy density in Eq.~\eqref{Teff-psi}. The  function $V_{\mathrm{MTY}}$ in Eq.~\eqref{eq:VMTY-series} differs from the effective potential in Eq.~\eqref{eq:Veff-series} beginning with the $(\psi^*\psi)^3$ term. The two-body  two-gradient interaction term in Eq.~\eqref{eq:Weff-NGK} does not appear in Eq.~\eqref{eq:WMTY-gradients}. Despite the many differences in their Lagrangians, the two effective field theories could be equivalent if they correspond to different definitions of the  complex field $\psi$.

%%%%%%%%%%%%%%%%%%%%%%%%%%%%
%%%%%%%%%%%%%%%%%%%%%%%%%%%%
%%%%%%%%%%%%%%%%%%%%%%%%%%%%
\subsection{$\bm{T}$-matrix elements}
\label{sec:Tmatch}

One way to show the equivalence of two different effective field theories is to verify that they give the same $T$-matrix elements. Equivalently, one can verify that they both reproduce the low-energy $T$-matrix elements of the same fundamental theory. In Ref.~\cite{Braaten:2016kzc}, the effective Lagrangian in Eq.~\eqref{Leff-psi} was constructed by requiring tree-level $T$-matrix elements to match those in the relativistic real scalar field theory  in the low-momentum limit. An error in the coefficient $v_4$ has been corrected in Ref.~\eqref{v4}. We will show that the $4 \to 4$ tree-level $T$-matrix element from the MTY effective Lagrangian in Eq.~\eqref{eq:LMTY_rescaled} matches that in the relativistic real scalar field theory   in the limit as the external 3-momenta go to 0 provided an interaction term with a time derivative is added to the MTY Lagrangian. With this correction, the first few tree-level $T$-matrix elements also match those from the effective Lagrangian in Eq.~\eqref{Leff-psi}. This suggests that the effective Lagrangian in Eq.~\eqref{Leff-psi}  and the MTY effective Lagrangian  in Eq.~\eqref{eq:LMTY_rescaled} are equivalent. The equivalence could be proven by showing that all the T-matrices are equal for all external momenta.

In  the relativistic real scalar field theory, the Feynman rule for a propagator with 4-momentum $p^\mu = (m+E, \bm{p})$ is $i/(p^2 - m^2 + i \epsilon)$. The Feynman rule for the $2n$ vertex is $-i \lambda_{2n}m^2/f^{n-2}$. In the MTY effective Lagrangian  in Eq.~\eqref{eq:LMTY_rescaled}, the complete propagator for a particle with energy $E$ and momentum $\bm{p}$ is $i/(E-\bm{p}^2/2m  + E^2/2m+ i \epsilon)$, which can be expressed more simply as $2im/(p^2-m^2+ i \epsilon)$, The momentum independent term in the Feynman rule for the $n \to n$ vertex, which comes from the potential $V_\mathrm{MTY} $ in Eq.~\eqref{eq:VMTY-series}, is $-i z_nm^2 f^2/(2mf^2)^n$. To obtain the $n\to n$ $T$-matrix element with the standard nonrelativistic normalization of single-particle states, the sum of diagrams  in the relativistic theory must be multiplied by a factor of $[4(m^2+\bm{p}^2)]^{-1/4}$ for  each external line with  momentum $\bm{p}$. The diagrams in the nonrelativistic effective field theory are the subset of diagrams in the relativistic theory in which the number of incoming and outgoing lines at every vertex are equal.
 
We begin by matching $2 \to 2$ $T$-matrix elements. The only tree-level diagram in the relativistic theory is the 4-particle vertex. The only tree-level diagram in the MTY effective theory is the $2 \to 2$ vertex. The $T$-matrix elements match in the low-momentum limit if the coefficient  in the $2\to 2$ vertex has the value $z_2 = \lambda_4$ in Eq.~\eqref{z2}.

%%%%%%%%%%%%%%%%%%%%%%%%%%%%%%%%%%%%%%%%%%%%%%%%
\begin{figure}[t]
\centerline{ \includegraphics*[width=12cm,clip=true]{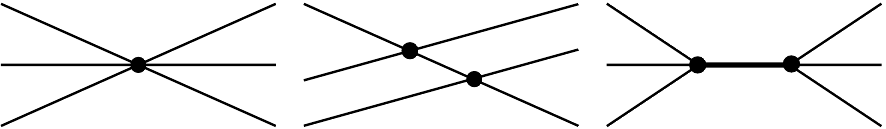} }
\vspace*{0.0cm}
\caption{The tree-level diagrams for low-energy $3 \to 3$ scattering in the relativistic  theory. The first 2 diagrams are also diagrams in a nonrelativistic effective theory. In the last diagram, the thicker line indicates a virtual particle whose invariant mass is approximately $3m$. 
}
\label{fig:3to3tree}
\end{figure}
%%%%%%%%%%%%%%%%%%%%%%%%%%%%%%%%%%%%%%%%%%%%%%%%

We proceed to match the $3 \to 3$  $T$-matrix elements. The tree-level diagrams for $3 \to 3$ scattering in the relativistic theory are shown in Fig.~\ref{fig:3to3tree}. The tree-level diagrams for $3 \to 3$  scattering in the MTY effective theory are the first 2 of the 3 diagrams in Fig.~\ref{fig:3to3tree}. The 2$^\mathrm{nd}$ diagram matches exactly in the two theories up to the normalization factors for external lines. The 1$^\mathrm{st}$ diagram in the MTY effective theory matches the sum of the 1$^\mathrm{st}$ and 3$^\mathrm{rd}$ diagrams in the  relativistic theory in the low-momentum limit if the coefficient  in the $3\to 3$ vertex has the value $z_3 =\lambda_6 +  \lambda_4^2/8$ in Eq.~\eqref{z3}.

%%%%%%%%%%%%%%%%%%%%%%%%%%%%%%%%%%%%%%%%%%%%%%%%
\begin{figure}[t]
\centerline{ \includegraphics*[width=16cm,clip=true]{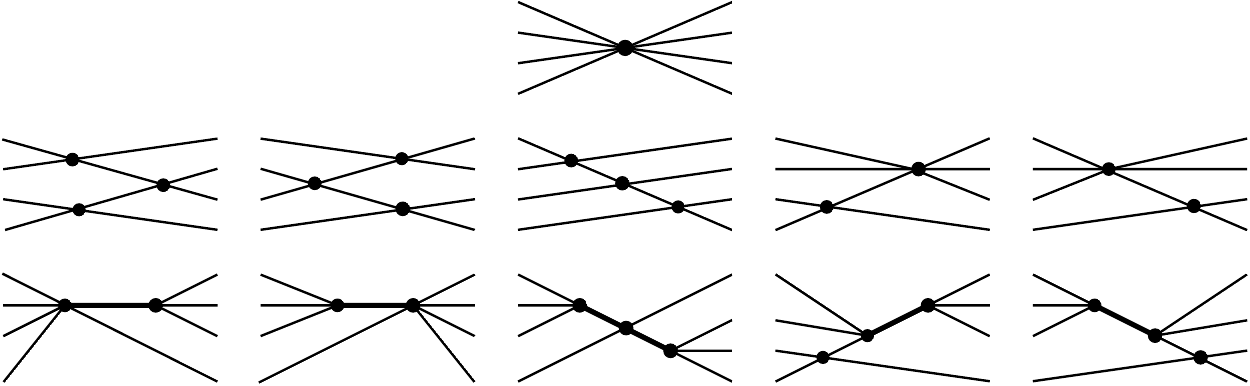} }
\vspace*{0.0cm}
\caption{The tree-level diagrams for low-energy $4 \to 4$ scattering in the relativistic theory. The first 6 diagrams are also diagrams in a nonrelativistic effective theory. In the last 5 diagrams, the thicker lines indicate virtual particles whose invariant mass is approximately $3m$. 
}
\label{fig:4to4tree}
\end{figure}
%%%%%%%%%%%%%%%%%%%%%%%%%%%%%%%%%%%%%%%%%%%%%%%%

Finally we match the $4 \to4$ $T$-matrix elements. The tree-level diagrams for $4 \to 4$ scattering in the relativistic theory are shown in Fig.~\ref{fig:4to4tree}. The tree-level diagrams for $4 \to 4$ scattering in the MTY effective theory are the first 6 of the 11 diagrams in  Fig.~\ref{fig:4to4tree}. The 2$^\mathrm{nd}$, 3$^\mathrm{rd}$, and 4$^\mathrm{th}$ diagrams match exactly in the two theories up to the normalization factors for external lines.  Given the value of $z_3$, the residues of the poles in the propagators in the 5$^\mathrm{th}$ and 6$^\mathrm{th}$ diagrams match. The sum of the 1$^\mathrm{st}$,  5$^\mathrm{th}$, and 6$^\mathrm{th}$ diagrams in the effective theory matches the sum of the 1$^\mathrm{st}$,  5$^\mathrm{th}$,  6$^\mathrm{th}$, and last five  diagrams in the relativistic theory in the low-momentum limit if the coefficient $z_4$ in the $4\to 4$ vertex has the value
%===========================
\begin{equation}
z_4 =    \lambda_8 +  \lambda_4 \lambda_6 - \frac{7}{8}\lambda_4^3.
\label{eq:z4-Tmatrix}
\end{equation}
%===========================
The coefficient of $\lambda_4 \lambda_6$ agrees with that in Eq.~\eqref{z4} if $x=1$, but the coefficient of $\lambda_4^3$ does not agree.

The source of the disagreement can be traced back to the failure of the MTY effective Lagrangian to reproduce correctly the momentum dependence of the $3 \to 3$ $T$-matrix element in the relativistic theory. In the relativistic theory, the $3 \to 3$ $T$-matrix element is the sum of the 3 diagrams in Fig.~\ref{fig:3to3tree}. The dependence on the 3-momenta  from the 2$^\mathrm{nd}$ diagram is correctly reproduced, but the dependence on the 3-momenta  from the 3$^\mathrm{rd}$ diagram is not. The dependence of the $3\to3$ vertex on the 3-momenta enters into the calculation of the $4 \to 4$ $T$-matrix element through the 5$^\mathrm{th}$ and 6$^\mathrm{th}$ diagrams in  Fig.~\ref{fig:4to4tree}. The problem can be fixed most easily by adding an interaction term with time derivatives to the MTY Lagrangian:\footnote{This time-derivative interaction term can be deduced from Eq.~(B.4) in an Appendix of Ref.~\cite{Mukaida:2016hwd} by interpreting the difference $\mu$ between the ``bare mass'' and the ``effective mass''as the differential operator $-i\partial_t/3$ acting on $\psi^3$.}
%===========================
\begin{equation}
\Delta W_\mathrm{MTY}=  
-\frac{1}{256}\lambda_4^2 \left(\frac{\psi^* \psi}{2 m f^2} \right)^2 \frac{i }{2}  \left( \psi^* \dot \psi - \dot \psi^*\psi \right)
+ \ldots.
\label{eq:WMTY-timederivs}
\end{equation}
%===========================
The contribution of this time-derivative  interaction term to the $3 \to 3$ $T$-matrix element for nonzero 3-momenta is comparable to that from the gradient interaction term in Eq.~\eqref{eq:WMTY-gradients}. This interaction term gives additional contributions to the $4 \to 4$ $T$-matrix element in the low-momentum limit from the 5$^\mathrm{th}$ and 6$^\mathrm{th}$ diagrams in  Fig.~\ref{fig:4to4tree}. When they are taken into account, matching with the relativistic theory gives the value of $z_4$ in Eq.~\eqref{z4} with $x=1$.

\subsection{Field redefinition}
\label{sec:Fieldredef}

Two different effective Lagrangians are equivalent if one can be transformed into the other by redefinitions of the fields. If the MTY effective Lagrangian in Eq.~\eqref{eq:LMTY_rescaled} is equivalent to the effective Lagrangian in Eq.~\eqref{Leff-psi}, it should be possible to transform it into  Eq.~\eqref{Leff-psi} by a redefinition of the complex field $\psi$.

The one-body terms in the effective Lagrangians in Eqs.~\eqref{Leff-psi} and \eqref{eq:LMTY_rescaled} both describe a particle with the relativistic energy-momentum relation $E= \sqrt{m^2+p^2}-m$. The  one-body terms  in Eq.~\eqref{eq:LMTY_rescaled} can be transformed into the one-body terms in Eqs.~\eqref{Leff-psi} and \eqref{Teff-psi} by a redefinition of the complex field $\psi$ that is nonlocal in space and time:
%===========================
\begin{equation}
\psi\longrightarrow
\left[1+\frac{1}{2m}\left(i\partial_t + \sqrt{m^2-\bm{\nabla}^2}-m\right)\right]^{-1/2}\psi.
\label{eq:field_redefinition-nonlocal}
\end{equation}
%===========================
This field redefinition can be expanded in powers of the operator in parenthesis 
and then expanded in powers of $\bm{\nabla}^2$.  Inserting the expansion into the 
terms $V_\mathrm{MTY}$ and $W_\mathrm{MTY}$ in the effective Lagrangian in Eq.~\eqref{eq:LMTY_rescaled} produces local  interaction terms that depend on time derivatives of $\psi$. To transform the effective Lagrangian into that in Eq.~\eqref{Leff-psi} would require subsequent field redefinitions that eliminate interaction terms with time derivatives without modifying the one-body terms. 

We will determine the field redefinitions required to eliminate those time-derivative interaction terms in the MTY effective Lagrangian that affect terms through order $(\psi^*\psi)^4$ in the effective  potential. For this purpose, we can ignore gradients of the field in the MTY Lagrangian. The expansion of the nonlocal field redefinition in Eq.~\eqref{eq:field_redefinition-nonlocal} to second order in the time derivative is
%===========================
\begin{equation}
\psi \longrightarrow \psi -(i/4m)\dot \psi-(3/32m^2)\ddot \psi.
\label{eq:field_redefinition-0}
\end{equation}
%===========================
After making this substitution in the MTY Lagrangian and adding a total time derivative that cancels the terms with a factor of $\ddot \psi$, the terms with no gradients  and up to two time derivatives are
%===========================-
\begin{subequations}
\begin{eqnarray}
\mathcal{L}_0  &=&
- V_\mathrm{MTY}(n),
\label{eq:L0NLO}
\\
\mathcal{L}_1 &=&
\Big[1 + \frac{1}{2m}V_\mathrm{MTY}'(n) \Big]\frac{i}{2}\left( \psi^* \dot \psi - \dot \psi^*\psi \right) ,
\label{eq:L1NLO}
\\
\mathcal{L}_2 &=&\frac{1}{32m^2} \big[ V_\mathrm{MTY}''(n) + 2 V_\mathrm{MTY}'(n)/n \big] 
\big( \psi^* \dot \psi - \dot \psi^*\psi \big)^2
\nonumber\\
&&\hspace{1cm}
-\frac{1}{32m^2} \big[ 3 V_\mathrm{MTY}''(n) + 2 V_\mathrm{MTY}'(n)/n \big] {\dot n}^2 .
\label{eq:L2NLO}\end{eqnarray}
\label{eq:L012NLO}%
\end{subequations}
%===========================
where $n = \psi^*\psi$ and $V_\mathrm{MTY}'$ and $V_\mathrm{MTY}''$ are derivatives of $V_\mathrm{MTY}(n)$ with respect to its argument.

The Lagrangian $\mathcal{L}_1$ in Eq.~\eqref{eq:L1NLO} includes a 2-body  interaction term with a time-derivative.  It  can be canceled by  a local field redefinition of the form 
%===========================
\begin{equation}
\psi \longrightarrow \psi + p_1 (n/2mf^2)  \psi .
\label{eq:field_redefinition-1}
\end{equation}
%===========================
This field redefinition changes the Lagrangians $\mathcal{L}_0$ and $\mathcal{L}_1$  in Eq.~\eqref{eq:L012NLO} into $\mathcal{L}_0'$ and $\mathcal{L}_1'$. The resulting effective potential is $-\mathcal{L}_0' =V_\mathrm{MTY}\big((1+P)^2 n \big)$, where $P(n)=p_1 (n/2mf^2)$. The 3-body coefficient $v_3$ defined by Eq.~\eqref{eq:Veff-series} is $v_3=z_3+36 p_1z_2$. The terms in $\mathcal{L}_1'$, including the 2-body interaction term, are
%===========================
\begin{equation}
\mathcal{L}_1' =
 \bigg[ 1+ \frac{z_2+16p_1}{8} \bigg( \frac{n}{2mf^2} \bigg)  \bigg]
 \frac{i}{2}\left( \psi^* \dot \psi - \dot \psi^*\psi \right) .
\label{eq:L1NLO'}
\end{equation}
%===========================
The coefficient required to cancel the 2-body interaction term with one time derivative is $p_1= - z_2/16$. The 3-body coefficient in the effective potential is determined to be 
%===========================
\begin{equation}
v_3 =  z_3 - \frac{9}{4} z_2^2.
\label{eq:v3-zn}
\end{equation}
%===========================
Upon inserting the results for $z_n$ in Eq.~\eqref{z234}, we reproduce the value of $v_3$ in Eq.~\eqref{v3}.

The  3-body  interaction term with one  time derivative and the  2-body  interaction terms with two time derivatives  can also be canceled by a more complicated local field redefinition of the form 
%===========================
\begin{equation}
\psi \longrightarrow \psi 
+ p_1 \bigg( \frac{n}{2mf^2} \bigg) \psi 
+ p_2 \bigg( \frac{n}{2mf^2} \bigg)^2 \psi 
+ \frac{q_1}{4m} \bigg( \frac{n}{2mf^2} \bigg)\,  i \dot \psi + \frac{r_0}{8m^2 f^2} \big(-i \psi^2 \dot \psi^*\big).
\label{eq:field_redefinition-2}
\end{equation}
%===========================
After inserting this field redefinition into the Lagrangian $\mathcal{L}_0' +\mathcal{L}_1'+\mathcal{L}_2'$, we expand to second order in time derivatives and we  add a total time derivative that cancels  the terms with a factor of $\ddot \psi$ or $\ddot \psi^*$. The resulting terms with 0, 1, and 2 time derivatives are $\mathcal{L}_0''$, $\mathcal{L}_1''$, and $\mathcal{L}_2''$. The expression for the effective potential $-\mathcal{L}_0''$ remains $V_\mathrm{MTY}\big((1+P)^2 n \big)$, except that now $P(n)=p_1 (n/2mf^2)+ p_2(n/2mf^2)^2$. The 4-body coefficient $v_4$ defined by Eq.~\eqref{eq:Veff-series} is $v_4=z_4+96 p_1 z_3+ 288 (2p_2+3p_1^2)z_2$.  The terms with one time derivative, including interaction terms up to  3-body, and the two-body terms with two time derivatives  are
%===========================
\begin{subequations}
\begin{eqnarray}
\mathcal{L}_1'' &=&
\bigg[ 1
+ \frac{z_2+16p_1}{8} \bigg( \frac{n}{2mf^2} \bigg)
\nonumber\\
&&\hspace{0.5cm}
+ \frac{z_3 + 6(4p_1-q_1-r_0)z_2 +48 (2p_2+p_1^2)}{48} \bigg( \frac{n}{2mf^2} \bigg)^2 \bigg]
 \frac{i}{2}\left( \psi^* \dot \psi - \dot \psi^*\psi \right),
\label{eq:L1NLO''}
\\
\mathcal{L}_2'' &=&\frac{3z_2-16(q_1+r_0)}{256m^2f^2} 
\big( \psi^* \dot \psi - \dot \psi^*\psi \big)^2
 - \frac{5z_2-16(q_1-r_0)}{256m^2f^2}
\, \dot n^2.
\label{eq:L2NLO''}
\end{eqnarray}
\label{eq:L12NLO''}%
\end{subequations}
%===========================
The coefficients required to cancel the 2-body interaction terms with two time derivatives are $q_1 =z_2/4$  and $r_0=- z_2/16$. The coefficient required to cancel the 3-body interaction term with one  time derivative is $p_2= -(16 z_3 - 39 z_2^2)/1536$. The 4-body coefficient in the effective potential is determined to be $v_4 =z_4-12z_2z_3 + 18z_2^3$. Upon inserting the results for $z_n$ in Eq.~\eqref{z234}, we obtain a  value for $v_4$ that differs from the one in Eq.~\eqref{v4}. The coefficient of $\lambda _4^3$ is 67/4 instead of 125/8.

The calculation above does not take into account the time-derivative interaction term in Eq.~\eqref{eq:WMTY-timederivs} that was omitted from the MTY effective Lagrangian. After the field redefinition in Eq.~\eqref{eq:field_redefinition-0}, it gives an additional term in the Lagrangian $\mathcal{L}_1$ in Eq.~\eqref{eq:L1NLO}, which results in an additional term in the Lagrangian $\mathcal{L}_1''$: %===========================
\begin{equation}
\Delta \mathcal{L}_1'' =
 \frac{\lambda_4^2}{256} \bigg( \frac{n}{2mf^2} \bigg)^2
 \frac{i}{2}\left( \psi^* \dot \psi - \dot \psi^*\psi \right) .
\label{eq:DeltaL2''NLO}
\end{equation}
%===========================
The coefficient $p_2$ required to cancel the 3-body interaction term with one  time derivative is now $p_2= -(16 z_3 - 39 z_2^2+3 \lambda_4^2)/1536$. The 4-body coefficient in the effective potential is determined to be  
%===========================
\begin{equation}
v_4 = z_4-12z_2z_3 + 18z_2^3 - \frac{9}{8} \lambda_4^2 z_2.
\label{eq:v4-zn}
\end{equation}
%===========================
Upon inserting the results for $z_n$ in Eq.~\eqref{z234} (with $x=1$), we reproduce the value for $v_4$ in Eq.~\eqref{v4}.

%%%%%%%%%%%%%%%%%%%%%%%%%%%%
%%%%%%%%%%%%%%%%%%%%%%%%%%%%
%%%%%%%%%%%%%%%%%%%%%%%%%%%%
\section{Generalized Ruffini-Bonazzola approach}\label{sec:ESW}

In Ref.~\cite{Ruffini:1969qy}, Ruffini and Bonazzola developed a method for finding solutions for Bose stars that consist of non-interacting bosons bound by their own gravitation field. Their method gave coupled  partial differential equations for the gravitational metric and the normalized wavefunction $\hat \psi_1(\bm{r})$ of a single-particle quantum state. Their equations can be obtained by making an ansatz for the real scalar quantum field $\phi\left({\bm r}, t\right)$  of the form
%===========================
\begin{equation}
\phi(\bm{r},t) = \frac{1}{\sqrt{2m}}
\left( \hat \psi_1(\bm{r})\, a_0\,  e^{-i  E_0 t} + 
 \hat \psi_1^*(\bm{r})\, a_0^\dagger\, e^{i  E_0 t}  \right),
\label{eq:phi_RB}
\end{equation}
%===========================
where $E_0$ is the energy of the ground state and $a_0$ is an operator that satisfies the commutation relation $[a_0 ,a_0^\dagger]=1$. The quantum state with $N$ bosons in the state created by $a_0^\dagger$ is $|N\rangle = (N!)^{-1/2}a_0^\dagger{}^N |0\rangle$, where the state  $|0\rangle$ satisfies  $a_0 |0\rangle= 0$ and $\langle 0 | 0 \rangle =1$. The partial differential equation for $\hat \psi_1(\bm{r})$ can be obtained by inserting the ansatz in Eq.~\eqref{eq:phi_RB} into the general-relativistic Klein-Gordon equation and taking the  matrix element between the $(N-1)$-boson state and the $N$-boson state in the limit $N\gg 1$. If instead of gravitational interactions, we consider a real scalar field with a self-interaction potential $V(\phi)$, the analogous equation is
%===========================
\begin{equation}
\big\langle N-1 \big| \big( \ddot \phi - \bm{\nabla}^2 \phi + m^2\phi + V'(\phi) \big) \big|N\big\rangle =0.
\label{eq:GRB1}
\end{equation}
%===========================
The resulting partial differential equation for $\hat \psi_1$ is  \cite{Eby:2014fya}
%===========================
\begin{equation}
{\bm \nabla}^2 \hat \psi_1+(E_0^2-m^2)\hat \psi_1-
2m V_\mathrm{naive}'(N\hat \psi_1^*\hat \psi_1)\, \hat \psi_1=0,
\label{eq:ESW-1}
\end{equation}
%===========================
where $V_\mathrm{naive}$ is the naive effective potential in Eq.~\eqref{eq:Veff-naive}.

In Ref.~\cite{Eby:2017teq}, Eby, Suranyi, and Wijewardhana developed a method for calculating corrections to  the  equations for  Bose stars that they referred to as a ``generalized Ruffini-Bonazzola approach''. They introduced an ansatz for  the real scalar quantum field that can be expressed in the form
%===========================
\begin{equation}
\phi(\bm{r},t) = \frac{1}{\sqrt{2m}} \sum_{n=0}^\infty
\left(  \hat \psi_{2n+1}(\bm{r}) \, a_0^{2n+1}\, e^{-i (2n+1) E_0 t}
+  \hat \psi^*_{2n+1}(\bm{r})\, (a_0^\dagger)^{2n+1}\, e^{+i (2n+1) E_0 t} \right).
\label{eq:theta-psi2n+1}
\end{equation}
%===========================
In Ref.~\cite{Eby:2017teq}, the Fourier modes  $\hat \psi_{2n+1}(\bm{r})$ were taken to be real-valued functions of a radial coordinate $r$ only, but they could in general be  complex functions of the position vector $\bm{r}$. They obtained  an arbitrarily large set of coupled nonlinear differential equations for $\hat \psi_{2n+1}(\bm{r})$ by inserting the ansatz in Eq.~\eqref{eq:theta-psi2n+1} into the  quantum field equation for $\phi(\bm{r},t)$ in Eq.~\eqref{eq:thetaEq} and taking matrix elements between the $(N-k)$-boson state and the $N$-boson state, where $k \ll  N$:
%===========================
\begin{equation}
\big\langle N-k \big| \big( \ddot \phi - \bm{\nabla}^2 \phi + m^2\phi + V'(\phi) \big) \big|N\big\rangle =0.
\label{eq:GRB}
\end{equation}
%===========================
If we keep only the $n=0$ term in the ansatz in Eq.~\eqref{eq:theta-psi2n+1}, the single equation in Eq.~\eqref{eq:GRB} with $k=1$ is Eq.~\eqref{eq:ESW-1}.

In Ref.~\cite{Eby:2017teq}, the authors devised a truncation scheme for the infinite set of coupled equations for the functions $\hat \psi_{2n+1}(\bm{r})$ in Eq.~\eqref{eq:theta-psi2n+1} that corresponds to an expansion in powers of $\Delta = \sqrt{m^2-E_0^2}/m$. Their coupled equations for $\hat \psi_{2n+1}(\bm{r})$  can be used to derive a self-contained differential equation for $\hat \psi_1(\bm{r})$ without increasing the parametric error. It is convenient to define the function $\psi_1(\bm{r})= N^{1/2}\,  \hat \psi_i(\bm{r})$, so that $\psi_1^*\psi_1$ is a number density. Eby et al.\ applied their truncation scheme at  order $\Delta^3$ to the scalar field theory whose potential  $V(\phi)$ has only a $\phi^4$ term. The truncation scheme gives coupled  partial differential equations for $ \psi_1$ and  $\hat\psi_3$. The equation for $\hat\psi_3$  was reduced  without increasing the parametric error to an algebraic  equation for $\hat\psi_3$ in terms of $\psi_1$. After eliminating $\hat\psi_3$ from the equation  for $ \psi_1$, it reduces to a self-contained partial differential equation:
%===========================
\begin{equation} 
\bm{\nabla}^2\psi_1 =
(m^2 - E_0^2) \psi_1   + \frac{\lambda_4 m^2}{2} \left(\frac{\psi_1^*\psi_1}{2mf^2} \right) \psi_1 .
\label{eq:order-psi1LO}
\end{equation}
%===========================
The interaction term at this order is essentially trivial, since it can be obtained from the $(\psi^*\psi)^2$ interaction term in the naive effective potential  in Eq.~\eqref{eq:Veff-naive}. 

In Ref.~\cite{Eby:2017teq}, the truncation method was applied at order $\Delta^5$ to the sine-Gordon model, whose interaction potential $V(\phi)$ is given by Eq.~\eqref{eq:V-sineGordon}. The potential $V(\phi)$ was truncated after the $\phi^6$ term. The truncation scheme of Ref.~\cite{Eby:2017teq} gives coupled partial differential equations for $ \psi_1$, $\hat\psi_3$, and  $\hat\psi_5$. The equations for $\hat\psi_3$ and $\hat\psi_5$  were reduced  without increasing the parametric error to algebraic  equations for $\hat\psi_3$ and $\hat\psi_5$ in terms of $\psi_1$ and its spatial derivatives. After eliminating $\hat\psi_3$ and $\hat\psi_5$ from the equation  for $ \psi_1$, it reduces to a self-contained partial differential equation:
%===========================
\begin{equation}
\bm{\nabla}^2\psi_1 =
(m^2 - E_0^2) \psi_1   - \frac{m^2}{2} \left(\frac{\psi_1^*\psi_1}{2mf^2} \right) \psi_1 
+ \frac{3m^2}{32} \left(\frac{\psi_1^*\psi_1}{2mf^2} \right)^2 \psi_1.
\label{eq:order-psi1NLO}
\end{equation}
%===========================

The equation of motion for $\psi$ derived from the MTY effective Lagrangian in Eq.~\eqref{eq:LMTY_rescaled}, including the potential in Eq.~\eqref {eq:VMTY-series} but excluding the gradient interaction terms in $W_\mathrm{MTY}$, is
%===========================
\begin{equation}
i\dot{\psi}-\frac{1}{2m}\ddot{\psi}=-\frac{1}{2m}{\bm \nabla}^2\psi
+ \frac{m}{2} \Bigg[ \frac{z_2}{2}\left(\frac{\psi^*\psi}{2mf^2} \right)
+ \frac{z_3}{12} \left(\frac{\psi^*\psi}{2mf^2} \right)^2 + \ldots \bigg] \psi.
\label{eq:MTYequation_motion}
\end{equation} 
%===========================
For the sine-Gordon model, the first two dimensionless coupling constants in Eq.~\eqref{eq:Veff-naive} are $\lambda_4 = -1$ and $\lambda_6=+1$, so the coefficients in Eqs.~\eqref{z2} and \eqref{z3} are $z_2=-1$ and $z_3=9/8$. We can obtain Eq.~\eqref{eq:order-psi1NLO} for $\psi_1(\bm{r})$ by substituting an ansatz with harmonic time dependence into Eq.~\eqref{eq:MTYequation_motion}:
%===========================
\begin{equation}
 \psi({\bm r},t) = \psi_1(\bm{r})\,  e^{-i (E_0-m) t}.
\label{eq:psi-psi_1}
\end{equation}
%===========================
Thus the improved equation for $\psi_1(\bm{r})$ in the sine-Gordon model in Eq.~\eqref{eq:order-psi1NLO} is consistent with the MTY effective Lagrangian with the effective potential truncated after the  $(\psi^*\psi)^3$ term. 

The authors of Ref.~\cite{Eby:2017teq} interpreted their truncation scheme at successively higher orders in $\Delta$ as defining a solution to the quantum field equation for $\phi$. However their ansatz in Eq.~\eqref{eq:theta-psi2n+1} was not derived from the quantum field theory. Whether the generalized Ruffini-Bonnazzola approach introduced in Ref.~\cite{Eby:2017teq} continues to give correct results at  higher orders is an open question. A good test of the method would be whether the next improved equation at order $\Delta^7$ is  consistent with the MTY Lagrangian with the effective potential truncated after the $(\psi^*\psi)^4$ term.

%%%%%%%%%%%%%%%%%%%%%%%%%%%%
%%%%%%%%%%%%%%%%%%%%%%%%%%%%
%%%%%%%%%%%%%%%%%%%%%%%%%%%%
\section{Summary and Discussion}
\label{sec:Summary}

A field theory with a real Lorentz-scalar field $\phi(x)$ can be described by a Lagrangian that is second order in both the time derivative and the spatial derivatives. If we consider only classical nonrelativistic field configurations in which gradients of $\phi$  are small compared to the mass $m$ and in which frequencies differ from the fundamental angular frequency $m$ by amounts much less than $m$, the field theory can alternatively be described by a classical nonrelativistic effective field theory (CNREFT) with a complex scalar field $\psi(x)$ whose Lagrangian is local and first order in the time derivative. Contrary to the claim in Ref.~\cite{Schiappacasse:2017ham}, the CNREFT does not necessarily require any limitation to low number density $\psi^*\psi$. This point is reinforced by the discovery by Namjoo, Guth and Kaiser of an exact canonical transformation between the real field $\phi(x)$ and the complex field $\psi(x)$ \cite{Namjoo:2017nia}. The nonrelativistic effective field theory may be able to describe field configurations with high number density as long as they have only long wavelengths and only frequencies close to $m$. When the number density $n = \psi^*\psi$ is high, approximating the effective potential by a truncated power series in $\psi^*\psi$ may  be insufficient. It may be necessary to determine  the effective potential at large values of $\psi^*\psi$ more accurately using a resummation method, such as that proposed in Ref.~\cite{Braaten:2016kzc}.

A CNREFT for a real Lorentz-scalar field was first explicitly constructed by us in Ref.~\cite{Braaten:2016kzc} using the matching methods of effective field theory. We considered the most general potential $V(\phi)$ with a $Z_2$ symmetry and a minimum at $\phi=0$. The real part $V_\mathrm{eff}$ of the effective potential has the expansion in powers of $\psi^*\psi$ in Eq.~\eqref{eq:Veff-series}. The first few coefficients in the expansion are given in Eqs.~\eqref{v234}. In Ref.~\cite{Braaten:2016kzc}, the coefficients were calculated through 5$^\mathrm{th}$ order in $\psi^*\psi$, but an error was made in the coefficients of the $(\psi^*\psi)^4$ and $(\psi^*\psi)^5$ terms. In the calculation of these coefficients, it is necessary to take into account interaction terms with gradients of $\psi$ and $\psi^*$ in the effective Lagrangian  \cite{HYZ:1801}. The correct coefficient $v_4$ of the $(\psi^*\psi)^4$ term is given in Eq.~\eqref{v4}.

Namjoo et al.\  used their exact transformation between $\phi(x)$ and $\psi(x)$ to verify the first nontrivial term in $V_\mathrm{eff}$, which is  the $(\psi^*\psi)^3$ term  \cite{Namjoo:2017nia}. They also determined a 2-body interaction term with gradients of $\psi$, which is given in Eq.~\eqref{eq:Weff-NGK}.

Mukaida, Takimoto and Yamada have  developed  a CNREFT by integrating out relativistic fluctuations of the real scalar field  \cite{Mukaida:2016hwd}. They considered a potential $V(\phi)$ with a minimum at $\phi=0$ and with a $\phi^3$ term as well as even powers of $\phi$. After the $\phi^3$ term is set to 0, the MTY effective Lagrangian differs from that in Ref.~\cite{Braaten:2016kzc} in many ways, including having a term that is second order in the time derivative of $\psi$ and having an effective potential that differs beginning at order $(\psi^*\psi)^3$. Mukaida et al.\  determined the $(\psi^*\psi)^2$ and $(\psi^*\psi)^3$ terms in the effective potential and part of the coefficient of the $(\psi^*\psi)^4$ term. They also determined a 3-body interaction term with gradients of $\psi$, which is given in Eq.~\eqref{eq:WMTY-gradients}. The 3-body interaction term with one time derivative  in Eq.~\eqref{eq:WMTY-timederivs} is equally important, but they failed to include this term in their effective Lagrangian.

We demonstrated the equivalence between our CNREFT in Ref.~\cite{Braaten:2016kzc} (with the corrected coefficient $v_4$) and the CNREFT of Ref.~\cite{Mukaida:2016hwd} (with the time-derivative term in Eq.~\eqref{eq:WMTY-timederivs} added to the effective Lagrangian). We demonstrated the equivalence in two different ways for the case of a potential $V(\phi)$ without a $\phi^3$ term. We showed that they both give the same  $T$-matrix elements for $2 \to 2$, $3 \to 3$, and $4 \to 4$ scattering in the low-momentum limit as the relativistic real scalar field theory. We also constructed a redefinition of the complex field that changes terms in the MTY effective Lagrangian, including the time-derivative interaction term in Eq.~\eqref{eq:WMTY-timederivs}, into the corresponding terms in the effective Lagrangian in Ref.~\cite{Braaten:2016kzc} with the corrected coefficient $v_4$. The nonlocal field redefinition in Eq.~\eqref{eq:field_redefinition-nonlocal} makes the one-body terms the same, but it introduces time-derivative interaction terms. The subsequent local field redefinition in Eq.~\eqref{eq:field_redefinition-2} eliminates the 2-body and 3-body interaction terms with one time derivative and no gradients and the 2-body interaction terms with two time derivatives and no gradients. They change the coefficients $z_3$ and $z_4$ in the MTY effective potential in Eq.~\eqref{z234} into the coefficients $v_3$ and $v_4$ in the effective potential in Eq.~\eqref{v234}.
 
In Ref.~\cite{Eby:2017teq}, Eby, Suranyi, and Wijewardhana developed a generalized Ruffini-Bonazzola approach that gives a sequence of improved equations for a complex field $\psi(x)$ with harmonic time dependence. They determined the first nontrivial interaction term in the specific case of the sine-Gordon model. We verified that their equation  at this order can be derived from the MTY effective Lagrangian with the $(\phi^*\phi)^3$ term in the effective potential. A nontrivial test of their formalism is whether their improved equation at the next order can be derived from the MTY effective Lagrangian with the $(\phi^*\phi)^4$ term in the effective potential.

In Ref.~\cite{Deng:2018xsk}, Deng et al.\ used  canonical transformations on the creation and annihilation operators of a real scalar field $\phi(x)$ with a $\phi^4$ interaction potential to derive an effective Hamiltonian for a complex field $\psi(x)$. They determined the coefficient $v_2$ of the $(\psi^*\psi)^2$ term in the effective potential, which is given in Eq.~\eqref{v2}. It is straightforward to extend their analysis to obtain the coefficient $v_3$ of the $(\psi^*\psi)^3$ term in Eq.~\eqref{v3} in the case $\lambda_6=0$. It would be interesting to see whether their method can also give the correct coefficient $v_4$ of the $(\psi^*\psi)^4$ term in Eq.~\eqref{v4}.

In Ref.~\cite{Visinelli:2017ooc}, Visinelli et al.\ pointed out that higher harmonics of the fundamental frequency of the real scalar field $\phi(\bm{r},t)$ could be important in the classical solutions for dense axion stars and for oscillons. They studied the effects of higher harmonics  on  oscillons in the sine-Gordon model by using first-order perturbation theory  to obtain coupled equations for the fundamental harmonic and the 3$^\textrm{rd}$ harmonic. They did not recognize that a self-contained equation for the fundamental harmonic could be obtained at second order in perturbation theory.  They also did not recognize that the problem of taking into account higher harmonics of the real scalar field  was solved in the CNREFT framework in Ref.~\cite{Braaten:2016kzc}.  The effects of the  higher harmonics are taken into account through the coefficients in the effective Lagrangian. The 3$^\textrm{rd}$ diagram for the $3 \to 3$ $T$-matrix element   in Fig.~\ref{fig:3to3tree} can be interpreted as a contribution from the 3$^\textrm{rd}$ harmonic of the real scalar field. Its effects are taken into account systematically through the coefficient $v_3$ of the $(\psi^* \psi)^3$ term in the effective potential  and through coefficients of  3-body gradient interaction terms.

The effective Lagrangian for CNREFT has a $U(1)$ symmetry, but the associated particle number is not conserved because there are  imaginary terms in the effective Hamiltonian. These imaginary terms take into account the loss of nonrelativistic particles from reactions in the relativistic theory that decrease the number of nonrelativistic particles and produce relativistic particles. The imaginary part $X_\mathrm{eff}$ of the effective potential has the expansion in powers of $(\psi^*\psi)^2$ in Eq.~\eqref{eq:Xeff-series}. The coefficients of the $(\psi^*\psi)^4$ and $(\psi^*\psi)^6$ terms were calculated in Ref.~\cite{Braaten:2016dlp}. They both vanish in the case of the sine-Gordon model. The cancellation of these coefficients suggests that $X_\mathrm{eff}$ may be zero to all orders in the sine-Gordon model.  If $X_\mathrm{eff}=0$, the loss rate of nonrelativistic particles in the sine-Gordon model must come from interaction terms in $Y_\mathrm{eff}$, which depend on gradients of $\psi$.  Calculations of those  interaction terms are in progress \cite{HYZ:1801}.

\begin{acknowledgments}
This research was supported in part by the Department of Energy under the grant DE-SC0011726 and by the National Science Foundation under the grant PHY-1310862. We thank D.~Kaplan, M.~Savage, and Q.~Wang for helpful discussions.
\end{acknowledgments}

\end{document}